\documentclass[nofootinbib,amsfonts,prd,aps]{revtex4}
\usepackage{graphicx}

\usepackage[usenames]{xcolor}

\begin{document}

\newcommand{\cev}[1]{\reflectbox{\ensuremath{\vec{\reflectbox{\ensuremath{#1}}}}}}

\title{Casimir effect in a quantum space-time}

\author{Rodolfo Gambini$^{1}$, Javier Olmedo$^{1,2}$,
	Jorge Pullin$^{2}$}
\affiliation {
1. Instituto de F\'{\i}sica, Facultad de Ciencias, 
	Igu\'a 4225, esq. Mataojo, 11400 Montevideo, Uruguay. \\
	2. Department of Physics and Astronomy, Louisiana State University,
	Baton Rouge, LA 70803-4001}

\begin{abstract}
We apply quantum field theory in quantum space-time techniques to
study the Casimir effect for large spherical shells. As 
background we use the recently constructed exact quantum solution
for spherically symmetric vacuum space-time in loop quantum
gravity. All calculations are finite and one recovers the usual
results without the need of regularization or renormalization. This
is an example of how loop quantum gravity provides a natural
resolution to the infinities of quantum field theories. 
\end{abstract}
\maketitle

Quantum field theory has been developed over the years with a series of
techniques for dealing with the infinities that arise, namely regularization and
renormalization. On curved space-times, however, certain problems remain which
have hampered efforts, for instance, to compute the back reaction of black hole
evaporation due to Hawking radiation \cite{wald}. It has been a long held belief
that when a theory of quantum space-time is developed, the point of view on
these issues may shift and some of the problems solved. Loop quantum gravity has
been steadily developed and in recent years has been used to treat vacuum
spherically symmetric space-times. In particular the space of physical states of
the theory has been found in closed form \cite{sphericalprl,javier} for that
case. One can therefore study quantum field theories living on such a quantum
space-time. A first step has already been taken in computing Hawking radiation
in such an approach \cite{hawking}. A remarkable characteristic is that the
discreteness of the quantum space-time acts naturally to regulate the infinities
of the quantum field theory. It is therefore of interest to study how such new
finite quantum field theories address potentially measurable effects like the
Casimir effect \cite{dewitt, milton}. The latter is a striking phenomenon
stemming from the energy of the vacuum in quantum field theory. Generically
speaking it implies that when one creates a bounded region in space-time, forces
will arise on the boundaries due to the suppression of certain quantum modes
that the boundary conditions impose on the quantum fields. From a technical
point of view continuum quantum field theory requires additional ingredients in
order to deal with divergences, like regularization by point splitting and
renormalization, exponential cutoffs	or even the use of functional analysis
techniques like the Riemann zeta function for the computation of the infinite
(usually divergent) mode sums. Though these techniques give finite results
independent of the scheme of regularization (and renormalization) employed, we
still lack a description free of these drawbacks. The intention of this paper is
to address some of these questions and their consequences for a quantum field
theory in a quantum space-time. Since loop quantum gravity puts at our disposal
\cite{sphericalprl,javier} the exact quantum states corresponding to an
effective background that is spherically symmetric (generically containing a
black hole) we will study the Casimir effect between two spherical
shells. They will be embedded inside a bounded region (provided, for instance, by two auxiliary
spherical shells) such that it covers an extensive portion of the entire
space-time. Although we can carry out our study throughout the whole space-time,
to simplify calculations we will restrict the study to the interior of the
shells and assume they are far away from the black hole horizon that generically
is present in spherically symmetric vacuum space-times, so one can use a planar
approximation in the computation of the Casimir effect. We will consider a
quantum scalar field living in a quantum spacetime as discussed in
\cite{hawking}, but this time interacting with two neighboring shells where 
the field vanishes. The quantization of the field is done within the Fock representation
whereas the background is quantized via loop quantum gravity in spherical
symmetry. The background quantum space-time \cite{sphericalprl,javier} is
characterized by a vector $\vec{k}$ that corresponds to the valences of the
links of the spin network it is based on and the value of the mass at spatial
infinity $M$. The components of $\vec{k}$ are proportional to the values of the
areas of the spheres of symmetry intersected by the spin network $A_i=4\pi
\ell_{\rm Planck}^2 k_i$. If one considers states that are eigenstates with
precise values of $\vec{k},M$, as opposed to superpositions, the main effect of
considering a quantum field theory on quantum space-time treatment is to
discretize the equations for the scalar field (since $M$ is a continuous
parameter we strictly speaking consider a narrow superposition of values of
$M$). The field takes values at the vertices of the spin networks only, as it is
the usual treatment of scalar fields in loop quantum gravity. If one considers
quantum states that are in a superposition, there exist additional effects to
the ones discussed here, but it does not change the main conclusions, i.e. that
all quantities are well defined without infinities. We will assume that the
quantum space-time is such that it approximates well a continuum smooth
geometry. This places some constraints on the values of $\vec{k}$: i) one would
like them to grow monotonically to avoid coordinate singularities, and ii) to
have small differences between successive components to avoid large ``jumps'' in
the values of the areas of the spheres of symmetry. We will also for simplicity take the
changes in adjacent values of $\vec{k}$ along the spin network to yield a
uniform spacing in the radial coordinate. Since $r^2_i=\ell_{\rm Planck}^2 k_i$,
the difference between two successive values of the radial coordinate has to be
at least $\ell_{\rm Planck}^2/(2r)$. For instance, in the exterior of a black
hole one can choose a uniform spacing with the lowest possible separation
$\Delta=\ell_{\rm Planck}^2/(2M)$. Notice that the bound allows lattice spacings
that are considerably smaller than the Planck scale, even for small black holes.
This in turn implies that the discrete equations for the quantum fields are
extremely well approximated by equations in the continuum, except in the extreme
trans-Planckian ultraviolet region of the modes of the fields.

We will proceed with the computation of the Casimir effect. On the one hand, we
consider a the scalar field interacting with two spherical shells of radii $r_0$
and $r_0+L$. We demand that the scalar field vanish at the shells as is usual
for calculations of the Casimir effect. Besides, since we assume that we are in
the asymptotic region where the background quantum metric is flat, we require
$r_0\gg 2M$. Otherwise the contributions due to the Schwarszchild geometry must
be considered. For convenience, we will carry out all the calculations with
respect to an observer at rest in $r=r_0$. On the other hand, we must consider
contributions of the field outside the shells as well. To do this, we place two
auxiliary shells at $r_0+L_0$ and $r_0-L_1$, such that $L_0\gg L$. The value of
$L_1$ can be selected arbitrarily, covering a large portion of space-time, but
still in the asymptotic region so we can still make the approximation that the
gravitational potential in the wave equation vanishes.

We will adopt a mode decomposition for the scalar field of the form
$u_{n,\ell,m} = \exp(-i\omega
t)R_{\ell}(\omega,r_j)Y_{\ell,m}(\theta,\varphi)/(\sqrt{2\pi\omega}r_j)$, where
$Y_{\ell,m}(\theta,\varphi)$ are the standard spherical harmonics. Besides,
$r_j=r_0+j \Delta$ is the discrete radial coordinate, with $j$ a suitable
integer and where $\Delta$ is the separation of the vertices of the spin network
of the quantum space-time. Therefore, the number of vertices on each region is
given by the quotient of its length over the step. For instance, inside the slab
of length $L$ we have $N_L=L/\Delta$ vertices. 

The radial modes fulfill the difference equation,
\begin{equation}
\frac{R_{j+1}-2R_{j}+R_{j-1}}{\Delta^2}+\omega^2R_{j}-\frac{\ell\left(\ell+1\right)}{r_j^2}
	R_{j}=0,
\end{equation}
with Dirichlet boundary conditions, where we have neglected the potential due to
the curvature of the Schwarzschild space-time since we assume the shells are far
away from the horizon. In the continuum, the corresponding solutions are linear
combinations of Bessel functions, and the dispersion relation is not known in
closed form but involves (unbounded) discrete frequencies. On the lattice, the
radial functions will approximate very well the continuum limit. Then, the field
decomposition is well approximated by these modes as
\begin{equation}\label{eq:field-mode}
	\phi\left(t,r_j\right) = 
\sqrt{\frac{2\pi}{N_L\Delta}}\sum_{n=1}^{N_L-1}\sum_{\ell=0}^{\frac{2r_0}{\Delta}}
	\sum_{m=-\ell}^\ell\left[
	\frac{a_{n,\ell,m}e^{-i\omega_{n,\ell} t}}{\sqrt{2\pi \omega_{n,\ell}}}
	\frac{\sin\left(\frac{\pi n j \Delta}{N_L \Delta}\right)}
	{r_j}Y_{\ell,m}\left(\theta,\varphi\right)
	+\frac{a^\dagger_{n,\ell,m}e^{i\omega_{n,\ell} t}}{\sqrt{2\pi
			\omega_{n,\ell}}}
	\frac{\sin\left(\frac{\pi n j \Delta}{N_L
			\Delta}\right)}{r_j}Y^*_{\ell,m}\left(\theta,\varphi\right)\right],
\end{equation}
where the dispersion relation is in very good agreement with
\begin{equation}\label{eq:disp-rel}
	\omega_{n,\ell}^2\simeq\frac{4}{\Delta^2}\sin^2\left(\frac{\Delta
		k_n}{2}\right)+\frac{\ell(\ell+1)}{r_0^2},
\end{equation}
and $k_n=(n\pi)/(N_L\Delta)$. The prefactor $\sqrt{\frac{2\pi}{N_L\Delta}}$
implies the modes are normalized. Moreover, and for the sake of simplicity, we
have replaced Bessel functions by sinusoidal ones. For modes with small $\ell$
this is a good approximation, but a more elaborate calculation would be needed
for modes of high angular momentum. Preliminary numerical studies justify these
approximations. Let us notice that the sum in $n$ is finite due to the
finiteness of the slab. The sum in 	$\ell$ is truncated since the dominant
contribution in our calculations will be due to modes of angular momentum $\ell$
lower than $2r_0/\Delta$ (the maximum mode frequency on the lattice times the radius
$r_0$). This bound is due to the asymptotic behavior of the Bessel functions for
high angular momentum. It is not oscillatory anymore but exponential. Therefore
no eigenfunctions can be found compatible with the Dirichlet boundary
conditions. In consequence, we will disregard them in our calculations.

Now, to compute the force due to the Casimir effect we will need to evaluate the
integral of the expectation value of the $T_{00}$ component of the stress-energy
tensor of the field in the region between the shells and compute its
derivative with respect to the separation $L$ between them.  The
relevant component of the 
energy-momentum tensor in the continuum \cite{fulling} is given by 
\begin{equation}
T_{00} = \frac{1}{2}
\dot{\phi}^2+\frac{1}{2}\nabla\phi\cdot \nabla\phi=
\frac{1}{2}
\dot{\phi}^2-\frac{1}{2}\phi\nabla^2\phi+\frac{1}{2}\nabla\left(\phi\nabla\phi\right),
\end{equation}
where the dot indicates derivation with respect time, $\nabla
\phi=(\partial_r\phi) \,\hat e_r+{1 \over r}{(\partial_\theta}\phi) \,\hat e_\theta
+ {1 \over r\sin\theta}{(\partial_\varphi}\phi)\,\hat e_\varphi$ is the standard
gradient and $\nabla^2\phi=\partial_r^2\phi+\frac{\hat L^2}{r^2}\phi$
is the Laplace operator in terms of the square of the standard angular momentum
operator  $\hat L^2$. Explicitly, on the basis of spherical harmonics it
fulfills $\hat L^2 Y_{\ell,m} (\theta, \varphi ) = -\ell (\ell + 1 ) Y_{\ell,m}
(\theta, \varphi )$. If we integrate this energy density inside the shells, the
total divergence contributes at the boundary, and it can be neglected since the field
vanishes on the shells (in the discrete theory this might not be true but the
nonvanishing contributions are of the order of the step of the lattice and can be disregarded).
Besides, the equations of motion can be employed to replace
$\phi\nabla^2\phi=\phi\ddot\phi$. At the end of the day we just need to compute the
expectation value of $T_{00}=\frac{1}{2}
\dot{\phi}^2-\frac{1}{2}\phi\ddot\phi$ with respect to a vacuum
state compatible with the Dirichlet boundary conditions imposed by the shells.
The corresponding creation and annihilation operators satisfy
$[a_{n,\ell,m},a^\dagger_{n',\ell',m'}]=\delta_{n,n'}\delta_{\ell,\ell'}\delta_{m,m'}$. 
In particular, the vacuum state $\vert 0\rangle_L$ for the slab of width $L$ is
defined such that $a_{n,\ell,m}\vert 0\rangle_L=0$.

To compute the expectation value of this component of the stress-energy tensor
in this region we would need
to concentrate on the derivatives of the field in the Green's
function associated
with the slab $G_+^L(x,x')=\langle 0_L\vert \phi(x)\phi(x')\vert
0_L\rangle$. On the one hand, after some calculations,
\begin{equation}
G_+^L\left(x;x'\right)\simeq\frac{1}{L}\sum_{n=1}^{N_L-1}\sum_{\ell=0}^{\frac{2r_0}{\Delta}}\sum_{m=-\ell}^\ell
	\frac{ e^{-i\omega_{n,\ell} \left(t-t'\right)}}{\omega_{n,\ell}}\frac{\sin\left(k_nz\right)}{r_j}\frac{\sin\left(k_nz'\right)}{r_{j'}}Y_{\ell,m}\left(\theta,\varphi\right)
	Y^*_{\ell,m}\left(\theta',\varphi'\right)
\end{equation}
with $z=r-r_0$, and similarly for $z'$, and $\omega_{n,\ell}$ is given in (\ref{eq:disp-rel}). It gives a good approximation for the discrete Green's function. 

We can calculate the first contribution to the $T_{00}$ component of the stress-energy tensor. For it, we only need to evaluate the Green's function for coincident radial and angular coordinates (but different times), and employ the identity
\begin{equation}
\sum_{m=-\ell}^\ell Y_{\ell,m}\left(\theta,\varphi\right)
Y^*_{\ell,m}\left(\theta,\varphi\right)=\frac{2\ell+1}{4\pi}.
\end{equation}
We also replace the sum in $\ell$ by an integral as well as we only keep the dominant terms. 
On then obtains the density
\begin{eqnarray}\nonumber
	\langle 0_L\vert \dot{\phi}^2\vert 0_L\rangle &=&
	\left.  \frac{\partial^2}{\partial t\, \partial t'}
	G_+^L\right\vert_{x=x'}=\frac{1}{4\pi L
		r_0^2}\sum_{n=1}^{N_L-1}\frac{2 r_0^2}{3}\sin^2\left(k_n
	z\right)\left[\frac{8}{\Delta^3}+\frac{12}{\Delta^3}\sin^2\left(\frac{\Delta
		k_n}{2}\right)\right.\\
	&&\left.-\frac{8}{\Delta^3}\sin^3\left(\frac{\Delta
		k_n}{2}\right)+\frac{3}{\Delta^3}\sin^4\left(\frac{\Delta
		k_n}{2}\right)+\ldots\right].\label{eq:F1}\\\nonumber
\end{eqnarray}
In this expression we have taken $r_j=r_0$ for simplicity. Let us emphasize that these
sums can be explicitly computed and give finite results at any $z\in[0,L]$, contrary to the
usual situation in continuum quantum field theory (due to the divergences at the boundary).
We have also replaced $\sin(2 \pi z/ \Delta)=0$ and
$\cos(2 \pi z/ \Delta)=1$ anywhere, since $z=\Delta j$. 

For the remaining contributions in $T_{00}$, recalling that we have employed the (discrete) equations of motion of the field, it is straightforward to see that
\begin{equation}\label{eq:F2}
\left.\langle 0_L\vert -\phi\ddot{\phi}\vert 0_L\rangle\right\vert=-\frac{\partial^2}{\partial^2 t'} \left.G_+^L\right\vert_{x=x'}=\langle 0_L\vert \dot{\phi}^2\vert 0_L\rangle.
\end{equation}

With this in mind, we can compute the $T_{00}$-component of the stress-energy 
tensor in the region between the smallest shells,
\begin{equation}\label{eq:exp-energ-tensor}
	\langle 0_L\vert T_{00}\vert 0_L\rangle=\langle 0_L\vert \dot{\phi}^2\vert 0_L\rangle,
\end{equation}
by means of (\ref{eq:F1}) and (\ref{eq:F2}), though its explicit expression 
is rather lengthy. 

For the scalar field inside the region with separation $\tilde L_0=(L_0-L)$ the
same construction can be adopted without additional considerations, just
replacing in the densities (\ref{eq:F1}) and (\ref{eq:F2}) the width $L\to (L_0-L)$ and afterwards
 $z\to(z-L)$. About the slab of width $L_1$, since we are
interested in variations of the energy of the system with respect to $L$, this
contribution will be a constant and we will not give it explicitly. The
corresponding expectation values of the $T_{00}$ component of the stress-energy
tensor will be then $\langle 0_{\tilde L_0}\vert T_{00}\vert 0_{\tilde
	L_0}\rangle$ and $\langle 0_{L_1}\vert T_{00}\vert 0_{L_1}\rangle$,
respectively.

To obtain the Casimir force, we simply calculate (minus) the derivative of the
energy of the system with respect to $L$. In particular, in the limit in which
$L_0\gg L$ and for small $\Delta$, we get for the force per unit area,
\begin{eqnarray}\nonumber
	{\rm P}&=& -\frac{d}{dL}\left(\int^0_{-L_1} dz \langle 0_{L_1}\vert T_{00}\vert
		0_{L_1}\rangle+\int_0^L dz \langle 0_L\vert T_{00}\vert
			0_L\rangle +\int_L^{L_0} dz \langle 0_{\tilde L_0}\vert T_{00}\vert 0_{\tilde
		L_0}\rangle\right)\\
		&=&-\frac{\pi^2}{480L^4}+{\cal
		O}(L_0^{-1})+{\cal
		O}(\Delta).
\end{eqnarray}

We then obtain the exact result \cite{ozcan,fulling}. Notice that since we are
working in the asymptotic region, it coincides with the planar case as
well. Let us emphasize that for the improved stress-energy tensor
\cite{callancolemanjackiw,fulling}, since it differs from the canonical one (in
the continuum) by a divergence, the very same result is obtained (up to
corrections of the order of the step of the lattice, precisely the contributions
we are neglecting). For the sake of completeness, we have also carried out the
same calculation for the $s$-mode of the scalar field ($\ell=0$), obtaining the exact 
result found in the continuum for the Casimir effect in 1+1 dimensions. It is
remarkable that without the cutoff that arise naturally from the quantum 
discreteness of the space-time this sum of energies would be ill defined.

It is worth commenting that the configuration we have adopted here to deal with
the problem of the Casimir effect might not have been employed before. We 
obtain the force as the variation of the energy of a closed
system, without any (possibly artificial) subtraction of an hypothetical
external vacuum energy. In fact, we need to add, nor subtract, appropriately the
energy of both slabs, considering the system as a whole, and take variations of
the total energy in order to achieve a meaningful physical result. Therefore,
our example sheds light in the understanding of one of the most prominent and
confirmed predictions of quantum field theory.

It is known that the Casimir force for spheres and planes is finite, but it
diverges for more general situations. It would be interesting to see if quantum
field theory in quantum space-times yields finite results in those cases and
compare them with (future) experiment. Moreover, of great interest will be to extend
these results to address the trace anomaly, which seems eminently feasible with
the techniques we used. Also our study has application in black hole evaporation
\cite{candelas}, where the expectation value of the stress-energy tensor with
respect to a suitable vacuum state plays an essential role for the computation
of the backreation. Our preliminary calculations yield a finite polarization of
the vacuum, avoiding as well the divergences of the continuum theory.

We have shown that one can compute the Casimir effect using quantum field theory
in quantum space-time techniques on the exact quantum spacetime of spherically
symmetric vacuum gravity of loop quantum gravity. All calculations result finite
and no regularization nor renormalization is needed. Remarkably, we reproduce
the result of quantum field theory in curved classical space-time which require
regularization and renormalization. Some authors have speculated \cite{fulling}
that gravitational effects might address those issues and this paper suggests
that they may. Our calculation is an example in which quantum gravity
successfully deals with the singularities that arise in ordinary quantum field
theory without additional hypothesis about them and yield sensible physical
results.

This work was supported in part by grant NSF-PHY-1305000, funds of the
Hearne Institute for Theoretical Physics, CCT-LSU and Pedeciba.


\begin{references}
	\bibitem{wald} R. Wald, ``Quantum field theory in curved space-time
	and black hole thermodynamics'', University of Chicago Press,
	Chicago, (1994). 
	\bibitem{sphericalprl}  R.~Gambini and J.~Pullin,
	``Loop quantization of the Schwarzschild black hole,''
	Phys.\ Rev.\ Lett.\  {\bf 110}, no. 21, 211301 (2013)
	[arXiv:1302.5265 [gr-qc]].
	\bibitem{javier}    R.~Gambini, J.~Olmedo and J.~Pullin,
	``Quantum black holes in Loop Quantum Gravity,''
	Class.\ Quant.\ Grav.\  {\bf 31}, 095009 (2014)
	[arXiv:1310.5996 [gr-qc]].
	\bibitem{hawking}  R.~Gambini and J.~Pullin,
	``Hawking radiation from a spherical loop quantum gravity black hole,''
	Class.\ Quant.\ Grav.\  {\bf 31}, 115003 (2014)
	[arXiv:1312.3595 [gr-qc]].
	\bibitem{dewitt} B. DeWitt, Phys. Rep. 19, 295 (1975).
	\bibitem{milton} K. Milton, ``The Casimir effect'', World Scientific,
	Singapore (2001).
	\bibitem{callancolemanjackiw} C. Callan, S. Coleman, R. Jackiw, 
	Ann. Phys. 59, 4273 (1970).
	\bibitem{fulling}S. Fulling, ``Aspects of quantum field theory in
	curved space-time'', Cambridge University Press, Cambridge (1989/1996). 
	\bibitem{ozcan}  M.~Ozcan,
	``Scalar Casimir effect between two concentric spheres,''
	Int.\ J.\ Mod.\ Phys.\ A {\bf 27}, 1250082 (2012)
	[arXiv:1207.4183 [quant-ph]].
	\bibitem{candelas} P. Candelas, Phys. Rev. D {\bf 21}, 2125 (1980). 
\end{references}
\end{document}